\documentclass[aps,prl,pdflatex,showpacs,twocolumn]{revtex4}
\usepackage{graphicx}

\begin{document}

\title{Non-linear space-time dynamics of ultrashort wave-packets in water}

\author{A. Matijo\v{s}ius$^1$, J. Trull$^2$ and P. Di Trapani}

\affiliation{Istituto Nazionale di Fisica della Materia (INFM) and
Department of Physics and Mathematics, University of Insubria, Via
Valleggio 11, IT-22100 Como, Italy}

\author{A. Dubietis, R. Piskarskas, A. Varanavi\v{c}ius, and A.
Piskarskas}

\affiliation{Department of Quantum Electronics, Vilnius
University, Saul\.{e}tekio Ave. 9, bldg.3, LT-2040 Vilnius,
Lithuania}

\date{\today}

\begin{abstract}
We have monitored the space-time transformation of 150-fs pulse,
undergoing self-focusing and filamentation in water, by means of
the nonlinear gating technique. We have observed that pulse
splitting and subsequent recombination apply to axial temporal
intensity only, whereas space-integrated pulse profile preserves
its original shape.
\end{abstract}

\pacs{190.5530, 190.5940, 260.5950, 320.7100}

\maketitle Spatial and temporal transformations of wave packets
which undergo self-focusing in transparent media with Kerr
nonlinearity attract a great deal of interest from the point of
view both of fundamental and applied science. Self-focusing of
ultrashort light pulses with power well exceeding that for
continuous-wave (CW) beam collapse gives rise to a variety of
spatial, temporal and interrelated (spatio-temporal, ST) effects
and has been a topic of intense theoretical and experimental
research \cite{Chernev, Rothenberg, Luther, Trippenbach, Gaeta,
Ranka98, Diddams, Fibich}. Discovery of a long range propagation
(filamentation) of ultrashort laser pulses in air boosted an
interest in ST transformation through filamentation dynamics in
gasses \cite{Mlejnek2, Couairon02a}, solids \cite{Tzortzakis01a}
and liquids \cite{Dubietis, Liu}.

Numerical models of different complexity have been elaborated to
study temporal, and, more generally, ST dynamics. Propagation of
intense light pulses in media with rather distinct optical
properties (amount of normal GVD, nonlinearity, etc; namely
gasses, solids and liquids) under various initial conditions
(pulse duration, beam width, power, wavelength) has been examined.
In spite of different regimes, temporal pulse splitting emerged as
a common feature resulting from the interplay between
self-focusing, self-phase-modulation and normal dispersion. The
open question is then what happens after the pulse splitting
occurs. A scenario leading to subsequent multiple splitting has
been predicted, but not observed directly (see  the discussion in
Ref~\cite{Berge02} and references therein). In Ref.~\cite{Diddams}
pulse recombination after splitting has been observed and
attributed to possible underlined multiple splitting.

More recently, two interpretations have been proposed for
understanding filament formation in condensed matter. Some of the
present Authors, after having experimentally demonstrated that
filaments cannot be treated as a self guided beams, have outlined
the key role plaid by non-linear losses in ruling a spontaneous
transformation from gaussian to conical (Bessel-like) wave, in the
frame of a CW model \cite{DubietisPRL}. In this context, the
conical wave provides the beam with a large (and not absorbed)
power reservoir that keeps "refuelling" the hot central spot and
so ensures stationarity, even in presence of non-linear losses.
Independently from this work, a second approach has been proposed
\cite{Kolesik} sharing with the first the genuine idea of
filaments as non-soliton, but conical waves. The important
difference is that here the Authors have outlined the key role of
chromatic dispersion, fully neglected in previous case, thus
interpreting the filament on the basis the dynamics of non-linear
X waves \cite{Valiulis, Conti, Di Trapani}.

The major problem concerning the experimental characterization of
the wave packet dynamics is that most of the measurements have
been limited either to the pure temporal domain, by on-axis auto
or cross correlation technique, or to the pure spatial one, by
time integrated CCD-based detection. An example of ST experimental
characterization has been that reported in Ref~\cite{Kumagai}. The
used technique, based on optical polarigraphy, has shown indeed a
quite poor resolution, and did not allowed the fine details of the
ST structure to be recovered. In a recent work related to the
investigation of the nonlinear dynamics of X-waves in quadratic
non-linear media some of the present authors have demonstrated a
very powerful, high ST resolution 3D-mapping technique, based on
the use of ultrafast $\chi^2$ (sum-frequency, SF) gate \cite{Di
Trapani, Trull}. The approach resembles the crosscorrelation
measurement to some extent, but instead of recording the
space-integrated (or simply on-axis) signal, here the entire
space-resolved SF profile is captured. The whole wave-packet is
then reconstructed from the assembly of time slices recorded at
different delay times.

In this work we apply the same technique to provide a complete
characterization of the nonlinear ST dynamics of ultrashort
wave-packets in water. The results confirm the occurrence of the
pulse recombination after splitting. However the feature of the
peripheral part of the beam seems to indicate that observed
recombination has to be linked not to multiple splitting, but in
contrast to refocusing of the non-trapped portion of the beam
\cite{DubietisPRL, Kolesik}.
\begin{figure}
\includegraphics{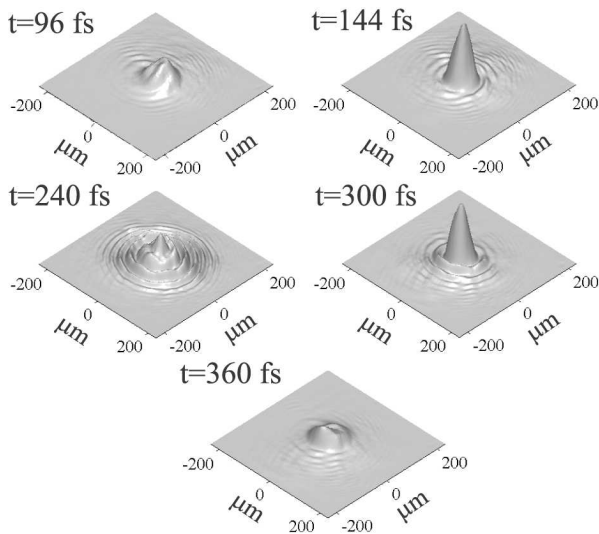} \caption{\label{fig:slices}
Intensity spatial profiles of different temporal slices of the
wave packet, I (x, y, t, z=22mm); t=0 at pulse leading edge.}
\end{figure}

\begin{figure}
\includegraphics{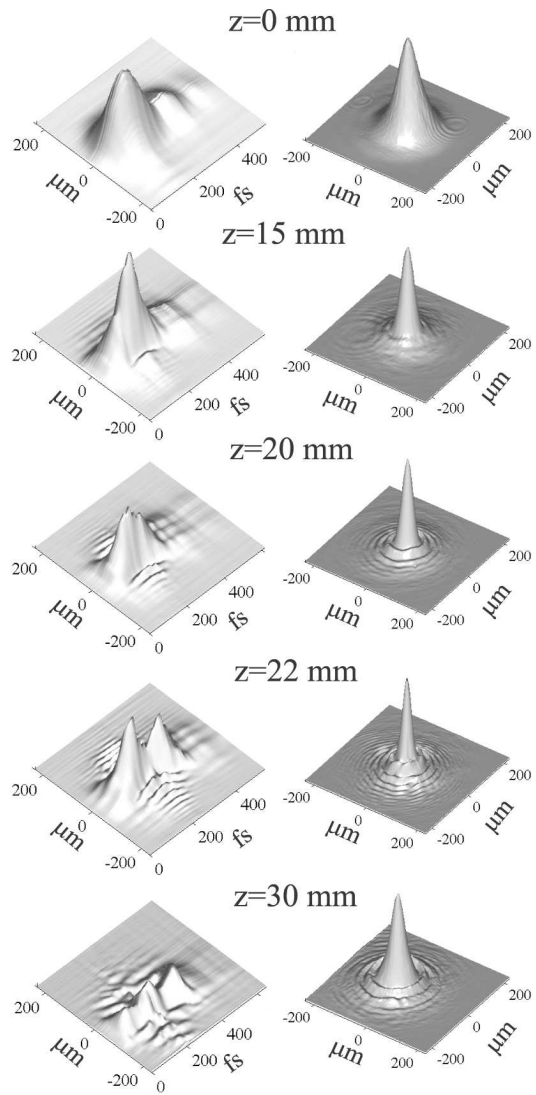} \caption{\label{fig:STevolution}
ST intesity profiles (left) and (time-integrated) normalized
fluence profiles (right) for different propagation lengths, z.}
\end{figure}

In our experiment, a light filament was excited by launching a
focused 0.9-$\mu$J ($P=5P_{cr}$, where $P_{cr}=1.15 MW$ is the
critical power for CW self-focusing), 150-fs, 527-nm pulses into a
syringe shaped water cell, using focusing geometry as described in
Ref.~\cite{Dubietis} (note that the actual focal length is 50cm
and not 50mm, as  written in Ref.~\cite{Dubietis} owing to typing
mistake). The input pulses were generated by a frequency
up-converted optical parametric amplifier (TOPAS, Light Conversion
Ltd.), pumped by 100-fs, 800-nm pulses delivered by Ti:Sapphire
laser system (Spitfire, Spectra Physics) at 1 kHz repetition rate.
A high contrast, 20-fs and 13-$\mu$J probe pulse with wavelength
centered at 710 nm was provided by a non-collinear optical
parametric amplifier (TOPAS White, Light Conversion Ltd.) pumped
by frequency doubled Ti:Sapphire laser pulses. The non-linear
gating was performed \emph{via} SF mixing the probe and the image
of the output wave-packet in a thin (20 $\mu$m) type I
phase-matching $\beta$-barium borate crystal under slightly
noncollinear geometry. SF signal centered at 302 nm was then
imaged onto high dynamic range (16-bit) CCD camera (Andor
Technology). We employed a probe pulse with constant intensity
over large area (note 1 mm FWHM diameter) and ensured that SF
process is performed in the low conversion limit. Then by spanning
the delay t using 12 fs steps (see Fig.~\ref{fig:slices}), the
complete intensity map I(x, y, t, z) was retrieved, the ultimate
temporal resolution being defined by the probe pulse width, by its
front steepness and by its intensity contrast.
\begin{figure}
\includegraphics{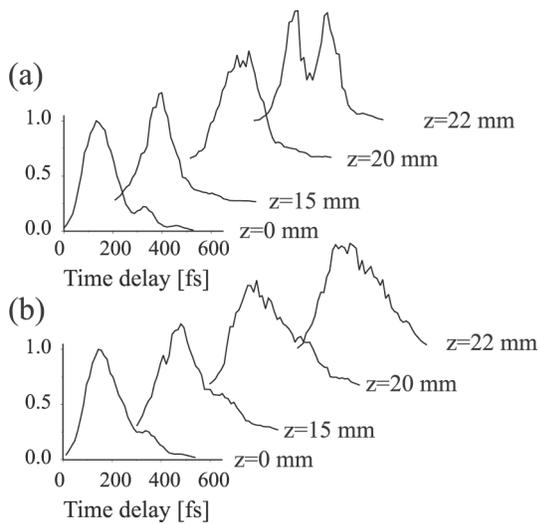} \caption{\label{fig:corr}
(a) Normalized intensity (temporal) profiles at the beam center,
I(x=0, y=0, t,  z) and (b) normalized power profiles, P(t, z), for
different propagation length, z.}
\end{figure}

Fig.~\ref{fig:STevolution} illustrates the full characterization
of the wave-packet at different z, as obtained by changing the
length of the water cell. In the right panel we show the
time-integrated energy density (fluence) distribution at the
output facet of the cell, as recorded by the CCD detector using
suitable imaging telescope. In the left panel we plot the
reconstructed ST intensity profiles (3D maps). We note that the
launched pulse has asymmetric temporal profile (see 3D map at z=0
mm in Fig.~\ref{fig:STevolution} with a shoulder on its trailing
edge). We expect that this shoulder, because of its low intensity,
does not play a major role in the nonlinear dynamics. The results
show that a single filament of $\sim45 \mu$m FWHM diameter is
formed, being surrounded by oscillating annular ring structure.
The ST maps show evident pulse splitting on axis at z=22 mm,
followed by pulse recombination at z=30mm. Out of the beam axis
intense modulated wings appear, whose intensity remain peaked at
the pulse central temporal coordinate. This last feature is
particularly important if one accounts that the major part of the
wave-packet energy is indeed contained into the outer part of the
beam. In fact, direct measurements by means of pinholes and
stoppers has shown that the filament contains only $20\%$ of the
pulse total energy \cite{DubietisPRL}. This feature is clarified
also by the results in Fig.~\ref{fig:corr}, where the temporal
distribution of the on-axis intensity and of the total power (the
space-integrated intensity) are compared. These results show that
the widely used term "pulse splitting" is, at least in this case,
incorrect. In fact, not the entire wave-packet gets split in the
time scale, but only the very central (on-axis) portion of it.
Indeed the power distribution remains peaked at the pulse center
at any value of z. These results let us foresee that pulse
recombination could appear because of refocusing of a portion of
the power distributed in these intense outer rings. To our
opinion, these findings provide an indication of the fact that the
wave-packet is attempting a reshaping toward a supported eigenmode
of the system, either with dominant Bessel type \cite{DubietisPRL}
or X-type \cite{Kolesik} profile, owing to the interplay among
self-focusing, dispersion and nonlinear losses.

In conclusion, by using high-resolution, high dynamic range
detection of the ST intensity profile, we have demonstrated that
pulse splitting and recombination of intense, ultra-short pulses
in water characterize just the temporal intensity profile \emph{at
the beam center}, but not the pulse as a whole. The entire
wave-packet dynamic seems ruled by relevant energy exchange
between the beam center and the self-built, slowly decaying,
oscillating tails.

The authors acknowledge valuable contribution of E.
Gai\v{z}auskas, G. Valiulis, G. Tamo\v{s}auskas, M. A. Porras and
the support from MIUR (Cofin01/FIRB01), DGI BFM2002-04369-C04-03
and EC CEBIOLA (ICA1-CT-2000-70027) contracts.

$^1$ Permanent address: Department of Quantum Electronics, Vilnius
University, Saul\.{e}tekio Ave. 9, bldg.3, LT-2040 Vilnius,
Lithuania.

$^2$ Permanent address: Departament de Fisica i Enginyeria
Nuclear, Universitat Politecnica de Catalunya, C/Colom 1, 08222
Terrassa, Spain.


\begin{references}

\bibitem{Chernev} P. Chernev, V. Petrov,  Opt. Lett. \textbf{17},
172 (1992).

\bibitem{Rothenberg} J. E. Rothenberg, Opt. Lett. \textbf{17}, 583
(1992); Opt. Lett. \textbf{17}, 1340 (1992).

\bibitem{Luther} G. G. Luther, A. C. Newell, J. V. Moloney, and E.
M. Wright, Opt. Lett. \textbf{19}, 789 (1994).

\bibitem{Trippenbach} M. Trippenbach and Y. B. Band, Phys. Rev. A \textbf{56},
4242 (1996).

\bibitem{Gaeta} A. L. Gaeta, Phys. Rev. Lett. \textbf{84}, 3582 (2000).

\bibitem{Ranka98} J. K. Ranka and A. L. Gaeta, Opt. Lett. \textbf{23}, 534 (1998).

\bibitem{Diddams} S. A. Diddams, H. K. Eaton, A. A. Zozulya, and
T. C. Clement, Opt. Lett. \textbf{23}, 379 (1998).

\bibitem{Fibich} G. Fibich, W. Ren, X.-P. Wang, Phys. Rev. E \textbf{67},
056603 (2003).

\bibitem{Mlejnek2} M. Mlejnek, E. M. Wright, and J. V. Moloney,
Phys. Rev. E \textbf{58}, 4903 (1998).

\bibitem{Couairon02a} A. Couairon, S. Tzortzakis, L. Berg\'{e}, M.
Franco, B. Prade, and A. Mysyrowicz, J. Opt. Soc. Am. B
\textbf{19}, 1117 (2002).

\bibitem{Tzortzakis01a} S. Tzortzakis, L. Sudrie, M. Franco, B.
Prade, A. Mysyrowicz, A. Couairon, and L. Berg\'{e}, Phys. Rev.
Lett. \textbf{87}, 213902 (2001).

\bibitem{Dubietis} A.~Dubietis, G.~Tamo{\v s}auskas, I.~Diomin, and
A.~Varanavi{\v c}ius, Opt.\ Lett.\ \textbf{28}, 1269 (2003).

\bibitem{Liu} W. Liu, S. L. Chin, O. Kosareva, I. S. Golubtsov,
and V. P. Kandidov, Opt. Commun. \textbf{225}, 193 (2003).

\bibitem{Berge02} L. Berg\'{e}, K. Germaschewski, R. Grauer, and J. J.
Rasmussen, Phys. Rev. Lett. \textbf{89}, 153902 (2002).

\bibitem{DubietisPRL} A. Dubietis, E. Gai\v{z}auskas, G. Tamo\v{s}auskas
and P. Di Trapani, arXiv:physics/0311072, submitted to Phys. Rev.
Lett. (7 July 2003).

\bibitem{Kolesik} M. Kolesik, E. M. Wright, and J. V. Moloney,
arXiv:physics/0311021.

\bibitem{Valiulis} G. Valiulis, J. Kilius, O. Jedrkiewicz, A. Bramati, S. Minardi,
C. Conti, S. Trillo, A. Piskarskas, and P. Di Trapani, in OSA
Trends in Optics and Photonics (TOPS), Vol. 57 OSA, 2001 (see
physics/0311081).

\bibitem{Conti} C. Conti, S. Trillo, P. Di Trapani, G. Valiulis,
A. Piskarskas, O. Jedrkiewicz, and J. Trull, Phys. Rev. Lett.
\textbf{90}, 170406 (2003).

\bibitem{Di Trapani} P. Di Trapani, G. Valiulis, A. Piskarskas, O.
Jedrkiewicz, J. Trull, C. Conti, and S. Trillo, Phys. Rev. Lett.
\textbf{91}, 093904 (2003).

\bibitem{Kumagai} H. Kumagai, S.-H. Cho, K. Ishikawa, K.
Midorikawa, M. Fujimoto, S. Aoshima, and Y. Tsuchiya, J. Opt. Soc.
Am. B \textbf{20}, 597 (2003).


\bibitem{Trull} J. Trull, O. Jedrkiewicz, P. Di Trapani, A. Matijo\v{s}ius, A.
Varanavi\v{c}ius, G. Valiulis, R. Danielius, E. Ku\v{c}inskas, A.
Piskarskas and S. Trillo, "Spatio-temporal 3D mapping of nonlinear
X-waves", Phys. Rev. E., in press.
\end{references}
\end{document}